\documentclass[12pt]{iopart}
\usepackage{graphicx}

\newcommand{\mean}[1]{\ensuremath{\left\langle #1 \right\rangle}}
\newcommand{\vars}[1]{\ensuremath{ \Delta^2 #1  }}
\newcommand{\Nbar}{\ensuremath{\bar{N}}}

\newcommand{\makesub}[1]{\textrm{\scriptsize{#1}}}

\newcommand{\lina}{\ensuremath{\bar{a}}}
\newcommand{\linad}{\ensuremath{\bar{a}^{\dagger}}}
\newcommand{\ad}{\ensuremath{a^\dagger}}
\newcommand{\bd}{\ensuremath{b^\dagger}}

\newcommand{\ai}{\ensuremath{a_{\textrm{\scriptsize{in}}}}}

\newcommand{\linai}{\ensuremath{\bar{a}_{\textrm{\scriptsize{in}}}}}

\newcommand{\alphai}{\ensuremath{\alpha_{\textrm{\scriptsize{in}}}}}

\newcommand{\XM}{\ensuremath{X_\textrm{\scriptsize{M}}}}
\newcommand{\PMM}{\ensuremath{P_\textrm{\scriptsize{M}}}}
\newcommand{\XL}{\ensuremath{X_\textrm{\scriptsize{M}}}}
\newcommand{\PLL}{\ensuremath{P_\textrm{\scriptsize{M}}}}

\begin{document}
\title[Optomechanical geometric phase]{Quantum state preparation of a mechanical resonator using an optomechanical geometric phase}
\author{K E Khosla$^1$, M R Vanner$^2$, W P Bowen$^1$, G J Milburn$^1$}

\address{$^1$ Center for Engineered Quantum Systems, University of Queensland, St Lucia 4072, Australia}
\address{$^2$Vienna Center for Quantum Science and Technology (VCQ) and 
Faculty of Physics, University of Vienna, Boltzmanngasse 5, A-1090 Vienna, Austria}
\ead{k.khosla@uq.edu.au}

\begin{abstract}
We demonstrate that a geometric phase, generated via a sequence of four optomechanical interactions, can be used to increase, or generate nonlinearities in the unitary evolution of a mechanical resonator. Interactions of this form lead to new mechanisms for preparing mechanical squeezed states, and preparation of non-classical states with significant Wigner negativity. 
\end{abstract}

\submitto{\NJP}

\section{Introduction}
The geometric phase is a phase imparted on the wavefunction of a quantum state by driving a system around a closed path in phase space \cite{Berry1987}. Within quantum optics this phase is widely used to create logic gates for quantum computing ~\cite{Sorensen2000,Milburn2000,Leibfried2003}, but has otherwise received little attention in optomechanics. Some notable exceptions come from recent proposals that have considered the effect of a geometric phase involving mesoscopic mechanical oscillators \cite{Vanner2012Planck,Vacanti2011}. In this paper we consider an optomechanical system \cite{Bose1999,Vanner2011,Marshall2003,MattEichenfield2009,Thompson2008,Mancini2002} with a time dependent optical drive that, via the optomechanical interactions, traverses a closed loop in phase space and thus imparts a geometric phase onto the mechanical element. 

It has been shown that time dependent control fields acting on an auxiliary system can be used to generate exotic quantum states \cite{Jacobs2007}. In an optomechanical system, the radiation pressure force due to light in an optical resonator can be used to accelerate a mechanical resonator. Driving the optical resonator (being used here as the control field) with a suitable sequence of laser pulses can be used to manipulate the motion of the mechanics. In our scheme, strong mechanical non-linearity is generated with a sequence of four pulsed optomechanical interactions in a measurement free process. During this sequence the optical field makes a closed loop in phase space and the mechanical oscillator obtains a phase proportional to the area enclosed within the loop \cite{Berry1987}, i.e. a Berry phase. It is shown how this phase produces a nonlinear potential for the mechanical resonator from the linear optomechanical  radiation pressure interaction, and in general increase existing mechanical nonlinearities. We then discuss how this mechanical nonlinearity can be used for quantum state preparation of the mechanical oscillator.

Our full protocol takes place within a small fraction of one period of the mechanical oscillator, and is hence robust against rethermalisation and decoherence - similar to ref \cite{Vanner2011}. During the state preparation the mechanical resonator remains stationary so only momentum quadrature is changed.

\section{Model}
The Hamiltonian for the optomechanical interaction is given by $H = \hbar g_0 \ad a \sqrt{2}(b + \bd)$, where $g_0$ is the interaction rate, $\hbar$ is the reduced Planck's constant, $a,~(\ad)$ and $b,~(\bd)$ are the annihilation (creation) operators of the optical and mechanical field respectively.

The $\sqrt{2}$ arises from our definition of $\XM = (b + \bd)/\sqrt{2}$. The equation of motion for $a$ is given by $\dot{a}(t) =  g_0 a(t) \XM - \kappa a(t) + \sqrt{2\kappa} \ai(t)$ where $\ai$ contains driving fields and vacuum noise. We now linearise about a coherent amplitude $\alpha(t)$ with quantum field operator $\lina(t)$, such that $a(t) = \alpha(t) + \lina(t)$. The coherent amplitude follows the equation of motion $\dot{\alpha} = -\kappa \alpha(t)  + \sqrt{2\kappa}\alphai(t)$ where $\alphai(t)$ the time dependent drive. The linearised field operator follows
\begin{eqnarray}
\dot{\lina}(t) = g_0 [\alpha(t) + \lina(t)]\XM - \kappa \lina(t)  + \sqrt{2\kappa} \linai(t).
\label{eq:lina}
\end{eqnarray}

The linearized approximation is made, and the $\lina(t) \XM$ term is dropped. This is valid even when $\alpha (t)$ is small as the thermal force, $F_{\mbox{\scriptsize{th}}}$, (e.g. in $\dot{\PMM} = g_0 (\alpha + \lina)\XM  + F_{\mbox{\scriptsize{th}}}$) on the oscillator will still be large compared to $g_0\lina \XM$. This gives an effective linearized Hamiltonian of the form,
\begin{eqnarray}
H /\hbar &=& g_0|\alpha(t)|^2\XM +  g_0 |\alpha(t)| (\rme^{\rmi \theta} \linad + \rme^{-\rmi\theta} \lina) \sqrt{2} \XM.
\label{eq:ham}
\end{eqnarray}

\noindent The first term generates a classical momentum imparted to the oscillator, $ \PMM  \rightarrow \PMM + g_0 \int \rmd t |\alpha(t)|^2 $, where the integral is over the duration of the pulse and therefore proportional to the input pulse area. As this term commutes with the Hamiltonian, we neglect it in the following discussion and only deal with the quantum optomechanical interaction. If we now consider a coherent pulse input at the cavity resonance frequency, the unitary for the quantum interaction in equation~\ref{eq:ham} is given by,
\begin{eqnarray}
U(\XL^\theta) = \exp\left[-\rmi \chi\XM\XL^\theta \right]
\label{eq:chidef}
\end{eqnarray}

\noindent where $\chi = \sqrt{2}g_0\int \rmd t |\alpha(t)|$ and $\XL^\theta = (\lina \rme^{-\rmi\theta} + \linad \rme^{\rmi\theta})/\sqrt{2}$. In calculating this, the mechanical period, $T_\textrm{\scriptsize{M}}$, is assumed to be large compared to the temporal width of the pulse $\sigma$, so the free evolution of the mechanics may be neglected. Consider a pulse interacting with the optomechanical system four times; the same pulse is recycled and undergoes four separate interactions. After each interaction the pulse is displaced in optical phase space such that the interaction Hamiltonian for each successive pulse is proportional to a sequence of quadrature phase operators of the cavity field  $\{\XL,\PLL,-\XL,-\PLL\}$, with $\XL = (\lina + \linad)/\sqrt{2}$, $\PLL = (\lina - \linad)/(\sqrt{2}\rmi)$. For each interaction, it is assumed that the pulse has the same temporal profile, this is valid as long as the pulse width is much larger than $\kappa^{-1}$. We may use the Baker-Campbell-Hausdorff formula \cite{Dynkin47} to show the effective unitary is given by,
\begin{eqnarray}
U_{\textrm{\scriptsize{eff}}} = U(-\PLL)U(-\XL)U(\PLL)U(\XL) = \exp[-\rmi\chi^2 \XM^2]
\label{eq:Udeff}
\end{eqnarray}

\noindent where $\chi^2$ is the area of the closed loop in the optical phase space. $U_{\makesub{eff}}$ can be seen as a geometric phase dependent on the position quadrature of the mechanical oscillator. This operator can be seen as a momentum displacement proportional to the mechanical position, $U^\dagger \PMM U = \PMM -\chi^2 \XM$. The momentum of the oscillator becomes correlated with its position, resulting in a squeezed quadrature in the final mechanical state.

\section{Experimental scheme}
\begin{figure}[hhht!!!]
\begin{center}
\includegraphics[width=140mm]{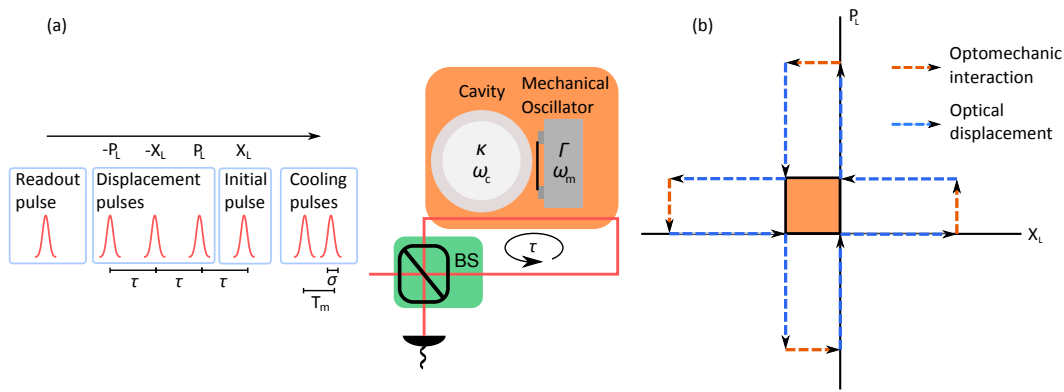}%
\caption{a) A schematic for an experimental protocol to realise mechanical nonlinearity via an optomechanical geometric phase. The required pulse sequence is shown entering the beam splitter. b) The evolution of the optical field in phase space. Blue lines show the coherent drive, with the orange lines representiong the optomechanical interaction. Each interaction gives a displacement in the quadrature orthogonal to the optical drive, with a magnitude depending on the position of the mechanical oscillator.}%
\label{fig:setup}%
\end{center}
\end{figure}

Figure~\ref{fig:setup} (a) shows a schematic of one possible experiment to realise the geometric phase. A coherent laser pulse of temporal width $\sigma$ enters a large fiber cavity with round trip time $\tau$ via a highly reflective beamsplitter. The pulse interacts with the mechanical oscillator via evanescent coupling from a toroidal cavity \cite{Anetsberger2009} with decay rate $\kappa$, and after exiting the toroidal cavity, is displaced in optical phase space. The optical displacement is performed using the highly reflective beam splitter, where a phase controlled laser pulse is used to displace away the coherent amplitude, and displace up in an orthogonal quadrature~\ref{fig:setup} (b). The pulse then repeats the optomechanical interactions and displacement three more times to give the four pulses sequence. 

The first pulse correlates the phase quadrature of the light ($\PLL$) with the position of the mechanics. The first optical displacement changes the coherent amplitude of the pulse from the $\XM$ ($\PLL$) to the $\PLL$ ($\XL$) quadrature, before the second optomechanical interaction. For the second interaction, $\XL$ is correlated with the mechanical position since it used to be $\PLL$ before the displacement. During the second interaction the back action of $\XL$ on the mechanical resonator effectively correlates the momentum of the oscillator with its position. As the position does not change over the these interactions interactions, correlating the momentum with the position produces a mechanical squeezed state. At this point the optical field is still correlated with the mechanical state, however, after the following two pulses the correlation is undone such that the final optical pulse is uncorrelated with the mechanical state, leaving the final state disentangled. The pulse sequence, including optomechanical interactions is shown in figure~\ref{fig:setup} (b).

To generate the geometric phase, the optomechanical system must satisfy the constraints, $T_\makesub{M}  \gg \tau > 4\sigma > 1/\kappa$. Setting $\tau > 4 \sigma$, ensures that each successive pulse decays out of the cavity prior to the next pulse entering, with only 0.01\% of the pulse remaining. Consequently interference between successive pulses can be neglected. Finally we require $T_\makesub{M}  \gg \tau$ to ensure the mechanics is near motionless during the four pulses to generate the geometric phase.

\begin{figure}[hhht!!!]
\begin{center}
\includegraphics[width=120mm]{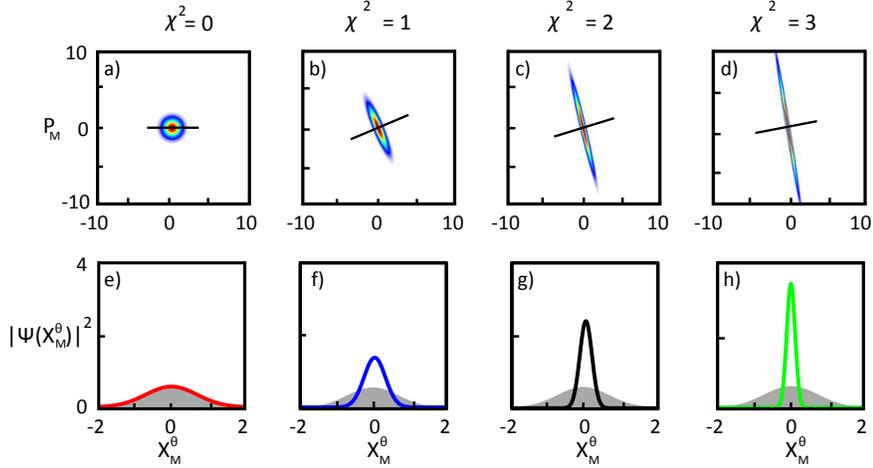}%
\caption{a) Effect of the unitary on the Wigner function of the mechanical oscillator, initially prepared in the ground state. The squeezed quadrature is marked by a line in each graph. This quadrature is maximally squeezed at an angle $\tan\theta = \sqrt{\chi^4 + 1} - \chi^2$ to the \XM~quadrature. b) The probability amplitude for the squeezed quadrature compared to the ground state value (shown in gray).}%
\label{fig:wigner}%
\end{center}
\end{figure}

Figure~\ref{fig:wigner} shows the effect of the four pulse sequence on the state of a mechanical oscillator initially prepared in the ground state, demonstrating how correlating $\XM$ and $\PMM$ leads to a squeezed quadrature from a ground state Wigner function. Increasing the nonlinearity $\chi$ benefits the protocol in two ways. Firstly it increases the effect of squeezing in the oscillator. Secondly it rotates the state so the squeezed quadrature aligns closer with the position quadrature, such that less time is required before the state can be verified (see section \ref{sec:Param}), and therefore the degradation in the squeezing due to thermalisation will be reduced.  

\section{Experimental parameters}
\label{sec:Param}
The previous section showed that under ideal conditions the geometric phase can be used to produce squeezed mechanical states. In this section we will consider experimental technicalities such as thermalisation of the mechanical oscillator, optical losses, and possible non-closing of the optical phase space loop.

Optical losses will have detrimental effects to this protocol. The classical attenuation from the beam splitter will result in the phase space loop remaining unclosed after the four pulse sequence. This can be corrected for by changing either the amplitude or phase of each displacement to counteract the attenuation. We must also consider the amplitude-noise back action on the momentum of the mechanics. In the absence of vacuum noise entering each cycle, any back action on the momentum in the $\XL~(\PLL)$ pulse will be reversed by the $-\XM~(-\PMM)$ pulse. However when vacuum noise is introduced at the beam splitters, the amplitude noise in the $-\XM$ pulse will no longer perfectly cancel the amplitude noise from the $\XM$ pulse. Unlike the attenuation of the classical amplitude this mechanism cannot be easily corrected for in the protocol. 

Since the beam splitters are 99:1 reflective, we expect $1-0.99^2 \approx 2\%$ vacuum noise to be imparted onto the oscillator from each of the $\XM,-\XL$ and $\PLL,-\PLL$ pulse pairs. If the loss over the total cycle (beamsplitter, fiber loss, input-output coupling etc - moddeled as an effective beamsplitter with vacuum input) was $\eta$, then $1-\eta^2$ vacuum noise would be introduced to the oscillator. The square arises from the fact the pulse must circulate twice before it cancels the noise, e.g. the noise imparted from the $\XM$ pulse will only be canceled two cycles later from the $-\XM$ pulse.

\begin{figure}%
\begin{center}
\includegraphics[width=85mm]{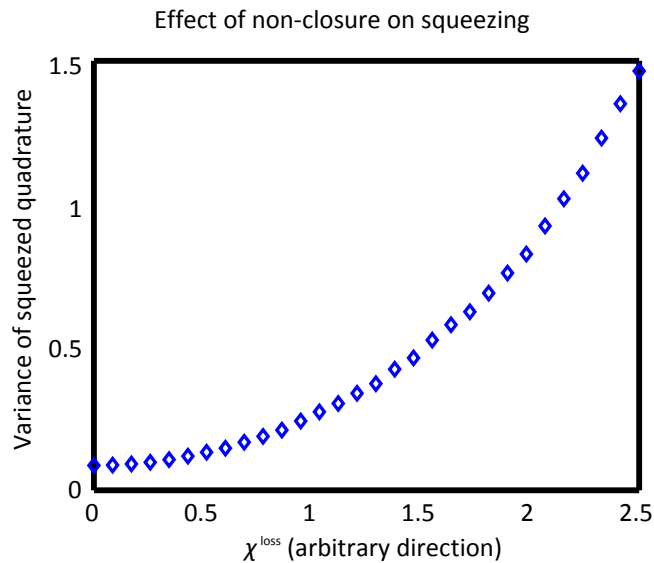}%
\caption{Effect of non-closure of the loop on the squeezed state for a fixed $\chi^2 = 1$. $\chi^{\makesub{loss}}$ is the magnitude of non-closure in an unknown quadrature $\XL^{\phi_{\makesub{loss}}}$. Once the variance of the squeezed state goes above one, squeezing becomes impossible as the amount of noise added from the non-closure of the loop is larger than the ground state variance.}%
\label{fig:nonclosed}%
\end{center}
\end{figure}

Even after correcting for the effect of losses, classical fluctuations in the pulse intensities use in the protocol could result in non-closure of the phase space loop. If the loop is not closed after the four pulse sequence, the effective unitary is given by $U = \exp[-\rmi\XM \sum_j \chi_j\XL^{\phi_j} - \rmi \chi^2 \XM^2]$, where $\chi^2 = \sum_{j=1,k>j}^{j = 4} \chi_j\chi_k [\XM^{\phi_k},\XM^{\phi_j}]/(2\rmi)$ and $\chi_i = 4g_0 \sqrt{N_i\sigma_i\sqrt{\pi/2}/\kappa}$ for a Gaussian pulse with temporal width $\sigma_i$ and $N_i$ photons. The second in the above unitary is the geometric phase. The first term entangles the light with the mechanics leaving the mechanical element in an entangeled state after the interaction. This can be viewed as a momentum displacement on the mechanics that depends on the optical field, $\mathcal{D}(-i \chi_{\makesub{loss}} \XL^{\phi_{\makesub{loss}}})$, where $\chi_{\makesub{loss}}$ is the displacement in the $\XL^{\phi_{\makesub{loss}}}$ quadrature that defines the final optical state. If $\chi_{\makesub{loss}}$ and $\phi_{\makesub{loss}}$ are unknown, this extra phase will reduce the observed squeezing. A homodyne measurement of the light lost from the beamsplitter will give an estimate of $\chi_{\makesub{loss}}$ and $\phi_{\makesub{loss}}$, meaning this error can be accounted for retrospectively. Figure~\ref{fig:nonclosed} shows how $\chi^{\makesub{loss}} \neq 0$ changes the squeezed state. Squeezing is possible for $\chi^{\makesub{loss}} < 2.1$, at this point $\vars{\XM} > 1$ meaning the noise added from the non-closure of the loop will degrade the observed squeezing no matter how squeezed the initial state is. 

Even if classical drifts in the optical displacements are corrected for, vacuum noise introduced by losses in the feedback loop can cause non-closure resulting in a mixed mechanical state. For a single pass efficiency in the fiber loop $\eta$, the cancellation of noise between pairs of displacement measurements (eg. $\XM$ and $-\XM$) will be degraded by a factor $1-\eta^2$, leading to a loop non-closure of (expression for nonclosure here). For realistic inefficiencies in the range of 10\%, (corisponding to $1-0.9^2 \approx 20\%$ vacuum noise) the loop nonclosure due to non-cancellation of noise is negligible $\chi_{loss}  = 4 g_0 \sqrt{0.2\sigma \sqrt{\pi/2}/\kappa} \ll 1$. Hence the squeezing is not significantly effected by the addition of vacuum noise. 

Here we consider the effect of thermalisation on the squeezed state. Thermalisation can have two detrimental effects. Firstly, phonon exchange with the bath during the four pulse sequence will render the dynamics over the pulse sequence non unitary and change the final mechanical state. Secondly phonons that enter during the time scale required for the squeezed quadrature to rotate into the measurable position quadrature will degrade the observed squeezing. The first of these effects can be neglected since the four puses can be very closely spaced with only a short delay between them. For example, for a mechanical oscillator with resonance frequency $\omega_\makesub{M} = 24$ kHz, and quality factor $Q = 10^5$, the pulse duration should be $\sigma  \simeq 10^{-8}$s, such that the time for four pulses (on the order of $10^{-7}$s), is much smaller than the time scale for one phonon to enter the oscillator, $1/(\Gamma\Nbar) \approx 10^{-5}$s at $1$K.

Consequently, only thermal phonons entering after the state has been prepared will be considered. Quantum Langevin equations were used to model the mechanical resonator coupled to a white noise thermal bath. The mechanical field was solved for the operators $b(t)$ and $\bd(t)$ to find the variance $\vars{\XM} = \mean{\XM^2} - \mean{\XM}^2$ 
\begin{eqnarray}
\vars{\XM} &=& \frac{\rme^{-2\gamma t}}{2}[\cos(2\omega t)(\mean{\XM^2}_0-\mean{\PMM^2}_0) + \sin(2\omega t)\mean{\XM\PMM + \PMM\XM}_0 \nonumber \\
& &+ \mean{\XM^2}_0 + \mean{\PMM^2}_0 -1] + (1-\rme^{-2\gamma t})(\Nbar + 1)
\end{eqnarray}

\noindent where $\mean{\XM^2}_0 = \Nbar+ \frac{1}{2}$,  $\mean{\PMM^2}_0 = (\Nbar+ \frac{1}{2})(1 + 4\chi^4)$ and $\mean{\XM\PMM + \PMM\XM}_0 = -4\chi^2 (\Nbar+ \frac{1}{2})$ are the expectation values after the geometric phase has been applied. To minimise the initial phonon number before the four pulse sequence we envisage a cooling via pulsed measurement as outlined in ref \cite{Vanner2011}. In this protocol, two pulses separated by 1/4 of the mechanical period are used to measure the oscillator in two orthogonal quadratures, leading to a low entropy state. The result in \cite{Vanner2011} shows an effective thermal phonon number of 
\begin{eqnarray}
\Nbar_{\makesub{eff}} \simeq \frac{1}{2}\left(\sqrt{1 + \frac{1}{\chi^4} + \frac{\pi \Nbar}{Q \chi^2}} - 1 \right)
\end{eqnarray}

\noindent where $\chi = 4g_0\sqrt{N_p\sigma \sqrt{\frac{\pi}{2}}/\kappa}$. This gives $\Nbar_{\makesub{eff}} \simeq 10$ for a $1$mm, $24$kHz SiN resonator with $Q\approx 10^5$ and photon number $N_\makesub{P}  = |\alpha|^2 \approx 10^6$. Using the SiN string mechanical oscillators considered in this paper, the effective phonon number is a achievable for resonance frequencies of $\omega_\makesub{M} < 70$kHz and length $L < 10$mm with a maximum incident photon flux of $\dot{N}_p = 10^{16}$Hz ($\approx 2$mW at 630nm). Although this is the initial phonon occupation, the bath occupation remains at $\Nbar \approx 10^5$.

SiN strings present a particularly attractive mechanical oscillator, high mechanical quality factors of up to $7\times 10^6$ have been observed, and the mechanical resonance frequency may be tuned via tensioning \cite{Verbridge2007,Schmid2011}. The protocol requires the mechanical period to be large compared to all other characteristic time scales. From this constraint we will limit the following analysis to low frequency, $\omega_m = 1-70$kHz, SiN strings. From \cite{Schmid2011}, the expected Q factor of a stressed SiN string of dimensions $L\times h \times w$ is
\begin{eqnarray}
Q = \left [ \frac{(n\pi)^2 E h^2}{12 S L^2} + 1.0887 \sqrt{\frac{E}{ S}}\frac{h}{L} \right]^{-1}Q_{\makesub{Bending}}
\end{eqnarray}
\noindent with $E = 241\pm 4$ GPa the Young's modulus of SiN, $Q_{\makesub{Bending}} = 17000$ the quality factor related to bending damping mechanisms, and $ S = 4\omega_m^2L^2\rho_\makesub{SiN}$ the tensile stress of SiN (with density $\rho_\makesub{SiN}$) in the high stress limit. 

The optomechanical coupling rate $g_0$ is calculated from evanescently coupled SiN string coupling rate $G = 200$ MHz/nm~\cite{Anetsberger2011} and the oscillators zero point motion: $g_0 = G x_0 = G \sqrt{\hbar/(2m\omega_\makesub{M})}$ with $m$ the effective mass of the mode. The pulse width and optical cavity decay rate are defined by $T_m = 10^{-3} \sigma = 5/\kappa$ to satisfy the experimental requirements.

\begin{figure}%
\begin{center}
\includegraphics[width=140mm]{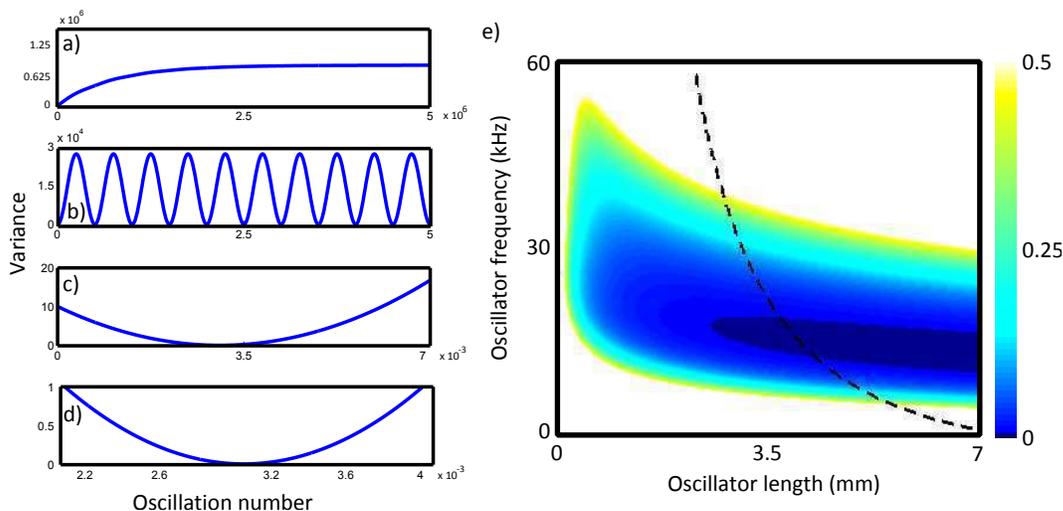}%
\caption{Progression of the variance of the mechanical oscillator over various time scales: a) Rethermalisation to the bath temperature. b) Oscillation of the variance from mechanical free evolution. c) Free evolution immediately after the four pulse sequence. d) Explicitly showing the squeezed region. e) Experimentally observable squeezing generated via a geometric phase of an oscillator initially cooled to $N_{\makesub{eff}} = 10$, averaged over the decay time of the cavity. The color axis gives the average variance in the position quadrature over one decay time of the oscillator. A value less than 0.5 indicates squeezing below the ground state variance. The graph color axis is truncated at 0.5.}%
\label{fig:results}%
\end{center}
\end{figure}

After preparing the mechanical quantum state with the four pulse sequence, it may be characterised with a measurement pulse. Any measurement of the variance of the position quadrature will be averaged over the optical cavity decay time. Hence we calculate the variance measurable over this time span as $\vars{\XM}^{\makesub{obs}} = \min_\tau\{\int_\tau^{\tau + 1/\kappa}\rmd t'\vars{\XM(t')} \}$, i.e. the average variance when the oscillator is maximally position squeezed. Figure~\ref{fig:results} (a)-(d) shows the temporal progression of the variance over various time scales after the four pulse sequence. Figure~\ref{fig:results} e) shows a plot of $\vars{\XM}^{\makesub{obs}}$ as a function of length and resonance frequency of a SiN string at $1K$ with cross section 157nm$\times 3\mu $m. These two parameters may be chosen independently by tensioning the string to the desired frequency. The dashed line on the figure represents the maximum experimental Q of $6.9\times 10^6$ \cite{Schmid2011} observed in a SiN string, with higher Q's required above the line. This figure shows it is experimentally feasible to achieve quantum squeezing for a wide range of geometries with the best squeezing of $\vars{\XM}\approx  10^{-2}$ predicted for a $3.5$mm long oscillator with $20$kHz resonance frequency and $Q = 5\times 10^6$. For all points in this figure, the initial state had an effective phonon number of $N_{\makesub{eff}} =10$ phonons; the maximum intracavity photon number to achieve such cooling was $10^6$ photons which is easily achievable. 

\begin{figure}[hhht!!!]
\begin{center}
\includegraphics[width=130mm]{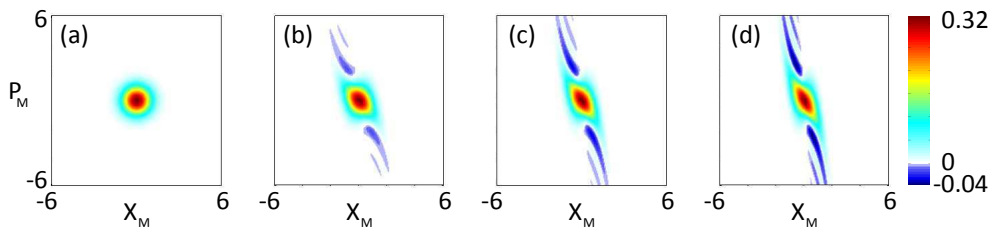}%
\caption{Wigner functions of the mechanical state of after a geometric phase interaction of a quadratically coupled mechanical oscillator for values of (a) $\chi^2 = 0$, (b) $\chi^2 = 0.066$, (c) $\chi^2 = 0.133$, (d) $\chi^2 = 0.2$. The momentum becomes correlated with the cube of the position - this can be seen in the Wigner function follows a profile proportional to $-x^3$ with negativity arising in the concave sections of the curve.}%
\label{fig:X4}%
\end{center}
\end{figure}

The $\XM^2$ appearing in the unitary is a result of the geometric phase changing a linear optomechanical potential interaction into an effective quadratic potential. If the mechanics was instead quadratically coupled to the light field \cite{Thompson2008,Jayich2008,Anetsberger2009,Purdy2010} ($H = g_0\hbar \ad a \XM^2$), the result would be a factor $\XM^4$ in the unitary - increasing the nonlinearity present in the Hamiltonian to forth order. In this case we may view the interaction as a position-cubed dependent displacement, $U =\exp [-\rmi\chi^2 \XM^4] = D(-\rmi\chi^2 x^3)$ correlating the momentum of the oscillator with the cube of its position. With significant quadratic coupling, this provides an avenue to generate quantum states of the oscillator involving significant Wigner negativity see figure \ref{fig:X4}. 

\section{Conclusion}
In summary, we have demonstrated how a geometric phase in an optomecnanical system can be used to generate a nonlinear unitary interaction from a linear Hamiltonian, and in general can increase the order of non-linearity present in a Hamiltonian. This provides a new tool in the optomechanics toolkit allowing mechanical squeezing and quantum state preparation. We have shown this method to be both robust and experimentally feasible.

\section{Acknowledgments}
This research was funded by the Australian Research Council Center of Excellence CE110001013.

\end{document}